\documentclass[aps,prl,letterpaper,superscriptaddress,twocolumn]{revtex4}

\usepackage{amstext,amsmath,amssymb,ulem,amsthm}
\pdfoutput=1
\usepackage{color}

\usepackage{times}
\usepackage{algorithm}
\floatname{algorithm}{Protocol}
\usepackage{ftnxtra}

\usepackage[T1]{fontenc}
\usepackage[utf8]{inputenc}
\usepackage{graphicx}
\usepackage[colorlinks]{hyperref}

\theoremstyle{plain}

\theoremstyle{definition}

\usepackage{float}

\newcommand{\beq}{\begin{equation}}
\newcommand{\eeq}{\end{equation}}

\newcommand{\ket} [1] {\vert #1 \rangle}
\newcommand{\bra} [1] {\langle #1 \vert}

\newcommand{\Tr}{\mathop{\mathrm{Tr}}}

\newcommand{\ba}{\begin{align}}
\newcommand{\ea}{\end{align}}
\newcommand{\bea}{\begin{eqnarray}}
\newcommand{\eea}{\end{eqnarray}}

\usepackage{tikz}	
\usepackage[hang,small,bf]{caption}  
\usetikzlibrary{backgrounds,fit,decorations.pathreplacing}  
\makeatletter

\@ifundefined{textcolor}{}
{%
 \definecolor{BLACK}{gray}{0}
 \definecolor{WHITE}{gray}{1}
 \definecolor{RED}{rgb}{1,0,0}
 \definecolor{GREEN}{rgb}{0,.4,0}
 \definecolor{BLUE}{rgb}{0,0,1}
 \definecolor{CYAN}{cmyk}{1,0,0,0}
 \definecolor{MAGENTA}{cmyk}{0,1,0,0}
 \definecolor{YELLOW}{cmyk}{0,0,1,0}
 }

\makeatother

\usepackage{bm}
\usepackage{mathtools}

\def\id{I}

\def\1{\mat{\id}}
\def\mat#1{\mathbf{#1}}

\renewcommand{\sout}[1]{}

\begin{document} 
\title{{Geometry of quantum correlations in space-time}}
\author{Zhikuan Zhao}
\thanks{These authors contributed equally to this work}
\affiliation{Singapore University of Technology and Design, 8 Somapah Road, Singapore 487372}
\affiliation{Centre for Quantum Technologies, National University of Singapore, 3 Science Drive 2, Singapore 117543}
\author{Robert Pisarczyk} 
\thanks{These authors contributed equally to this work}
\affiliation{Centre for Quantum Technologies, National University of Singapore, 3 Science Drive 2, Singapore 117543}
\affiliation{Mathematical Institute, University of Oxford, Oxford OX2 6GG, UK}
\author{Jayne Thompson}
\affiliation{Centre for Quantum Technologies, National University of Singapore, 3 Science Drive 2, Singapore 117543}
\author{Mile Gu}
\affiliation{Centre for Quantum Technologies, National University of Singapore, 3 Science Drive 2, Singapore 117543}
\affiliation{Complexity Institute, Nanyang Technological University,
18 Nanyang Drive, Singapore 637723, Singapore}
\affiliation{School of Mathematical and Physical Sciences, Nanyang Technological University, Singapore}
\author{Vlatko Vedral}
\affiliation{Centre for Quantum Technologies, National University of Singapore, 3 Science Drive 2, Singapore 117543}
\affiliation{Clarendon Laboratory, Department of Physics, University of Oxford, Parks Road, Oxford OX1 3PU, U.K.}
\affiliation{Department of Physics, National University of Singapore, 2 Science Drive 3, 117542, Singapore}
\author{Joseph F. Fitzsimons}
\email{joseph_fitzsimons@sutd.edu.sg}
\affiliation{Singapore University of Technology and Design, 8 Somapah Road, Singapore 487372}
\affiliation{Centre for Quantum Technologies, National University of Singapore, 3 Science Drive 2, Singapore 117543}

\begin{abstract}
The traditional formalism of non-relativistic quantum theory allows the state of a quantum system to extend across space, but only restricts it to a single instant in time, leading to distinction between theoretical treatments of spatial and temporal quantum correlations. 
Here we  unify the geometrical description of two-point quantum correlations in space-time. Our study presents the geometry of correlations between two sequential Pauli measurements on a single qubit undergoing an arbitrary quantum channel evolution together with two-qubit spatial correlations under a common framework. We establish a symmetric structure between quantum correlations in space and time. This symmetry is broken in the presence of non-unital channels, which further reveals a set of temporal correlations that are indistinguishable from correlations found in bipartite entangled states.
\end{abstract} 

\maketitle

The study of quantum correlations has given rise to valuable insights into fundamental physics, as well as promising prospects of quantum technologies \cite{bell1964einstein, ekert1991quantum, harrow2004superdense}. In the setting where observables are defined across spatially extended systems at a single time, substantial progress has been made in the context of entanglement \cite{horodecki2009quantum}. Particularly, the geometry of spatial correlations has been recognised as an important witness for entanglement in quantum systems \cite{schwinger1960geometry,Horodecki1996Information}. However the geometrical description of quantum correlations in the setting where observables are defined across different instances in space-time remains relatively under-explored.

In the usual formalism of quantum theory, the state of a system can extend across space, but is only defined at a particular instant in time. This distinction between the roles of space and time contrasts with relativity \cite{isham1993canonical} where they are treated in a more even-handed fashion, and has led to a preference to study temporal quantum correlations in a rather separated manner from their spatial counter-parts \cite{leggett1985quantum, brukner2004quantum, budroni2013bounding}. In order to study quantum correlations for observables defined across space-time, we make use of the pseudo-density matrix (PDM) formalism introduced in \cite{fitzsimons2015quantum} as an extended framework of quantum correlations, which generalises the notion of a quantum state to the temporal domain, treating space and time on an equal footing.

In this paper, we focus on analysing the simplest and most fundamental case, that of two-point correlation functions. In the spatial setting, this corresponds to bipartite quantum correlations, which can exhibit entanglement. In the temporal setting, we consider correlations between two sequential measurements on a single qubit. We work in the general framework where a qubit is free to undergo arbitrary quantum evolution between the measurements.
For the special case when the initial qubit is maximally-mixed, we show the resulting set of temporal correlations can be represented as a tetrahedron in the real space that is a reflection of its well-known spatial counterpart \cite{Horodecki1996Information}. We further identify the geometrical constraints on all components of a two-point PDM, hence completely classifying the geometry of two-point quantum correlations in space-time.

The density matrix of a quantum state can be viewed as a representation of the expectation values for all possible Pauli observables of a system.
This can be naturally extended into the temporal domain and used to define the PDM \cite{fitzsimons2015quantum} as
\begin{align}
R=\frac{1}{2^n}\sum\limits_{i_1=0}^{3}...\sum\limits_{i_n=0}^{3}\langle\{\sigma_{i_j}\}^n_{j=1}\rangle\bigotimes\limits_{j=1}^{n}\sigma_{i_j}, \label{eq:PDM}  
\end{align}
where $\sigma_0=\mathrm{I}$, $\sigma_1=\mathrm{X}$, $\sigma_2=\mathrm{Y}$ and $\sigma_3=\mathrm{Z}$. The sub-indices $j$ of each $i$ label different measurement events in the system. The factor $\langle\{\sigma_{i_j}\}^n_{j=1}\rangle$ denotes a correlation function of a size-$n$ sequence of Pauli measurements $\sigma_{i_j}\in\{\sigma_0,...,\sigma_3\}$. Physically, it is the expectation value of the product of the $n$ Pauli observables. Note that $R$ remains a Hermitian matrix with unit trace. Furthermore, if the measurement events are space-like separated, $R$ is positive semi-definite and hence is a valid density matrix. However, the structure of $R$ does not exclude the possible existence of negative eigenvalues. In the presence of negative eigenvalues, the Pauli measurements cannot be interpreted as having come from measurements on distinct sub-systems of a common quantum state. 
The novelty introduced by the PDM formalism is that local measurements can happen at arbitrary time instances, in contrast to the case for conventional density matrices. The presence of negative eigenvalues is a witness to this causal relationship, hence it is natural to quantify temporal correlations with the trace norm. A measure of causality was thus introduced as $f_{tr}(R)=\|R\|_{tr}-1$, which possesses desirable properties in close analogy with entanglement monotones for spatial correlations \cite{fitzsimons2015quantum}.  

Here we take a single-qubit system $\rho_A$ subject to a quantum channel between two measurement events at times $t_A$ and $t_B$. The channel is described by a completely positive trace-preserving (CPTP) map $\varepsilon_{B|A}$, which maps the family of operators from the state space $\mathcal{H}_A$ at $t_A$ to the state space $\mathcal{H}_B$ at $t_B$.
The PDM representation $R_{AB}$, of such a quantum system across $[t_A,t_B]$ is given by
\begin{align}
R_{AB}=	(\mathcal{I}_A\otimes \varepsilon_{B|A})\{\rho_A\otimes\frac{\mathrm{I}}{2},SWAP\}, \label{eq:PDM2}
\end{align} 
where $SWAP=\frac{1}{2}\sum_{i=0}^3\sigma_i\otimes\sigma_i$ and $\mathcal{I}_A$ denotes the identity super-operator acting on $A$ (see Appendix). 
This is in agreement with the Jordan product representation given in Ref. \cite{horsman2017can}:
\begin{align}
R_{AB}=\{\rho_A\otimes\frac{\mathrm{I}}{2},E_{AB}\}, \label{Jordan}
\end{align}
where $E_{AB}=\sum_{ij}\left(\mathcal{I}_A\otimes\varepsilon_{B|A}\right)\left(\ket{i}\bra{j}_A\otimes\ket{j}\bra{i}_B\right)$ is an operator
acting on $\mathcal{H}_A\otimes\mathcal{H}_B$ that is Jamio\l{}kowski-isomorphic to $\varepsilon_{B|A}$. The correlations described by $R_{AB}$ are "purely" temporal in the sense that the underlying dynamics are defined by a CPTP map on a single qubit.

\begin{figure}[h!]
\centering
\includegraphics[width=0.7\linewidth]{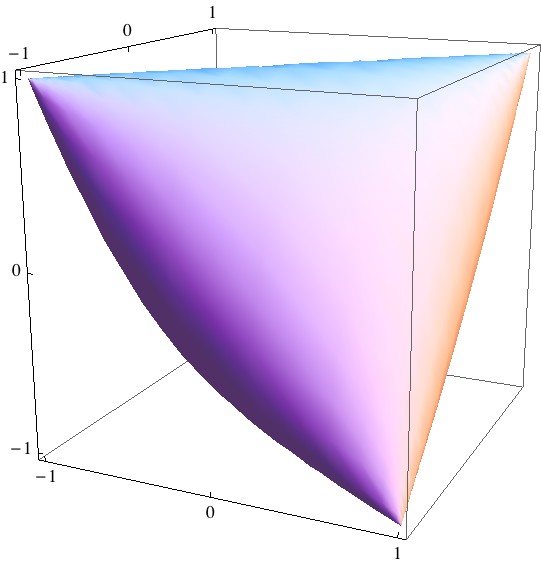}
\caption{The surface enclosing the set of possible values of two-point temporal correlations in the real space of $\{\left<\sigma_1\sigma_1\right>,\left<\sigma_2\sigma_2\right>,\left<\sigma_3\sigma_3\right>\}$. The equation of the surface is derived in the Appendix (see Eq. \eqref{trig}).}
\label{fig:inflate_tetrahedron}
\end{figure}

The set of three Pauli correlations $\langle\sigma_k\sigma_k\rangle=\Tr\left[(\sigma_k\otimes\sigma_k) R_{AB}\right]$ fully characterises any two-point correlations $\langle\sigma_k\sigma_l\rangle$ up to local unitary transformations, for $k,l=1,2,3$. We therefore illustrate the set of attainable $\langle\sigma_k\sigma_k\rangle$ as points in the real coordinator space $\{\left<\sigma_1\sigma_1\right>,\left<\sigma_2\sigma_2\right>,\left<\sigma_3\sigma_3\right>\}$ in FIG. \ref{fig:inflate_tetrahedron}, which depicts the geometry of two-time temporal Pauli correlations. The figure presents a parametric plot of the equations: $\left<\sigma_1\sigma_1\right>=\cos(u)$, $\left<\sigma_2\sigma_2\right>=\cos(v)$ and $\left<\sigma_3\sigma_3\right>=\cos(u-v)$, where $v\in[0,\pi], u\in[0,2\pi]$. It is worth noting that a similar structure was found in \cite{budroni2013bounding} in the context of Leggett-Garg inequalities, where the correlations among three sequential observables were considered.

We now focus on the cases where the initial system $\rho_A$ is maximally-mixed. The set of spatial correlations described by two-qubit density matrices has been studied in \cite{Horodecki1996Information}. It can be pictured in the space of $\{\left<\sigma_1\sigma_1\right>,\left<\sigma_2\sigma_2\right>,\left<\sigma_3\sigma_3\right>\}$  as the convex hull enclosed by the tetrahedron $\mathcal{T}_s$ with vertices of odd parity $(1,1,-1)$, $(1,-1,1)$, $(-1,1,1)$ and $(-1,-1,-1)$. These vertices correspond to the four Bell states. The set of temporal correlations described by $R_{AB}$ with $\rho_A=\frac{\mathrm{I}}{2}$ is the reflection of $\mathcal{T}_s$ in the $\left<\sigma_1\sigma_1\right>$-$\left<\sigma_3\sigma_3\right>$ plane. The resulting tetrahedron $\mathcal{T}_t$ has vertices of even parity $(1,-1,-1)$, $(1,1,1)$, $(-1,-1,1)$ and $(-1,1,-1)$.

\begin{figure}
\centering
\includegraphics[width=1.0\linewidth]{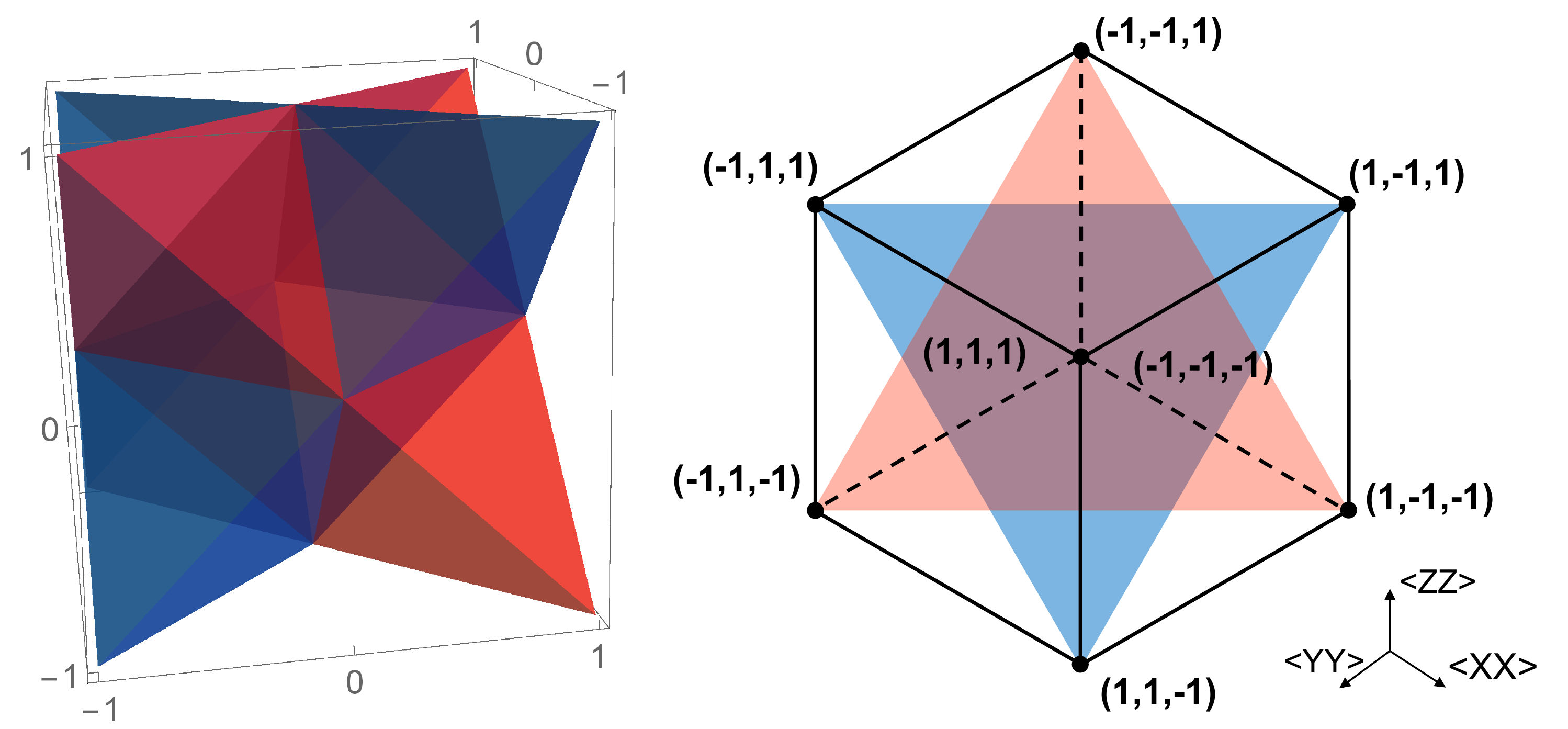}
\caption{\textbf{Left}: A 3-D visualisation of the spatial and temporal tetrahedrons with the blue region representing $\mathcal{T}_s$, and the red region representing $\mathcal{T}_t$.\\
\textbf{Right}: A perspective plot viewing from the $(-1,-1,-1)$ direction. The intersection of the spatial and temporal tetrahedra forms an octahedron that corresponds to the set of separable states. The purple hexagon is a projection of the octahedron overlap, while the blue and red triangles are projections of $\mathcal{T}_s$ and $\mathcal{T}_t$ respectively. The two tetrahedra are inscribed in the cube, which corresponds to the most general achievable region and is formed by a mixture of spatial and temporal correlations. It is clear that the cube is the largest possible set of space-time quantum correlations, since $-1\le\langle\sigma_A\sigma_A\rangle\le1$, and the set of possible correlation functions forms a convex set.}
\label{fig:StarInCube}
\end{figure}

The set $\mathcal{T}_t$ can derived from the relation $R_{AB}=\frac{1}{2}E_{AB}$ when $\rho_A=\frac{\mathrm{I}}{2}$. A partial transpose over sub-system $A$, which geometrically corresponds to the reflection, yields
\begin{align}
R^{\mathcal{PT}}_{AB} =& \left(\mathcal{I}_A\otimes\frac{\varepsilon_{B|A}}{2}\right)\sum\limits_{ij}\ket{ii}\bra{jj}_{AB}=\rho_{AB}^{\text{Choi}}(\varepsilon_{B|A}),\label{choi}
\end{align}
where $\rho_{AB}^{\text{Choi}}(\varepsilon_{B|A})$ is the Choi matrix of $\varepsilon_{B|A}$ \cite{choi1975completely}. For arbitrary choices of $\varepsilon_{B|A}$, the Choi matrices describe the same set of correlations, $\mathcal{T}_s$ as two-qubit density matrices. As the partial transpose over sub-system $A$ generates a reflection in the $\left<\sigma_1\sigma_1\right>$-$\left<\sigma_3\sigma_3\right>$ plane, the set $\mathcal{T}_t$ is simply an inverted copy of $\mathcal{T}_s$, see FIG. \ref{fig:StarInCube}. Interestingly, $T_t$ also describes temporal correlations for an arbitrary input state $\rho_A$  but with the channel restricted to be unital. The calculation leading to this observation is shown in the Appendix. As such, there is a conditional reflective symmetry between the sets of spatial and temporal quantum correlations. This symmetry is found to be broken in the presence of non-unital channels, giving rise to the set of temporal correlation shown in FIG. \ref{fig:inflate_tetrahedron}.

The Peres-Horodecki criterion \cite{horodecki1996separability} implies that the octahedron region formed by the overlap between the two tetrahedra $\mathcal{T}_t$ and $\mathcal{T}_s$ corresponds to the set of separable states, where both marginals are maximally-mixed. This insight allows us to make a natural connection between the entanglement measure, negativity $
f_{\mathcal{N}}(\rho_{AB})=\frac{1}{2}(\|\rho_{AB}^{\mathcal{PT}}\|_{tr}-1)$ \cite{vidal2002computable} and the causality
measure $f_{tr}$. Consider a two-qubit state $\rho^{\text{Choi}}_{AB}$ as the Choi matrix of $\varepsilon_{B|A}$ in Eq. \eqref{eq:PDM2}, leading to
$
f_{tr}(R_{AB})=2f_{\mathcal{N}}(\rho^{\text{Choi}}_{AB})
$.
The entanglement measure $f_{\mathcal{N}}$ can be visualised as the Euclidean distance $D_{s}$ between a point in $\mathcal{T}_s$ and the nearest point in the octahedron, such that $D_{s}=\frac{4f_{\mathcal{N}}}{\sqrt{3}}$ \cite{mundarain2007concurrence}. Hence, by analogy we can establish a geometrical interpretation for $f_{tr}$ as the Euclidean distance $D_t$ between a point in $T_t$ and the nearest point on the face of the octahedron, such that $D_t=\frac{2f_{tr}}{\sqrt{3}}$.

Beyond the geometry of the purely temporal and spatial correlations, a two-point PDM generally describes an arbitrary mixture of spatial and temporal correlations. Consider sequential Pauli measurements, $\sigma_A$ and $\sigma_B$ on one sub-system of a maximally-entangled pair. If the sub-system evolves through a CP-map, $\langle\sigma_A\sigma_B\rangle$ lies in the $\mathcal{T}_t$ as shown. However if a SWAP operation is applied before the second measurement, then the reduced dynamics on sub-system $A$ will no longer be described by a CP-map \cite{pechukas1994reduced}. Under these conditions the correlations $\langle\sigma_A\sigma_B\rangle$ will span $\mathcal{T}_s$. Furthermore, if SWAP is applied probabilistically, the possible correlations span the entire volume of the cube formed by the vertices of $\mathcal{T}_t$ and $\mathcal{T}_s$, fully inscribing the spatial and temporal tetrahedra. The geometry of various types of two-point correlations in space-time is depicted in FIG. \ref{fig:StarInCube}.

It is worth remarking that the "inflated tetrahedron" in FIG. \ref{fig:inflate_tetrahedron} inscribes a larger volume than $\mathcal{T}_t$, and therefore partially overlaps with the non-separable region in $\mathcal{T}_s$. Physically this implies that the correlations generated by spatial entanglement can be partially mimicked by temporal correlation described by a single-qubit PDM, and that it is impossible to distinguish between the two cases by only examining the correlation statistics. This can be contrasted with the vertices of $\mathcal{T}_s$ that correspond to maximally-entangled states. The inability to simulate correlations generated by Bell states with temporal measurements is linked to the impossibility of constructing a quantum universal-NOT gate \cite{buvzek1999optimal}.

\begin{figure}
\centering
\includegraphics[width=1\linewidth]{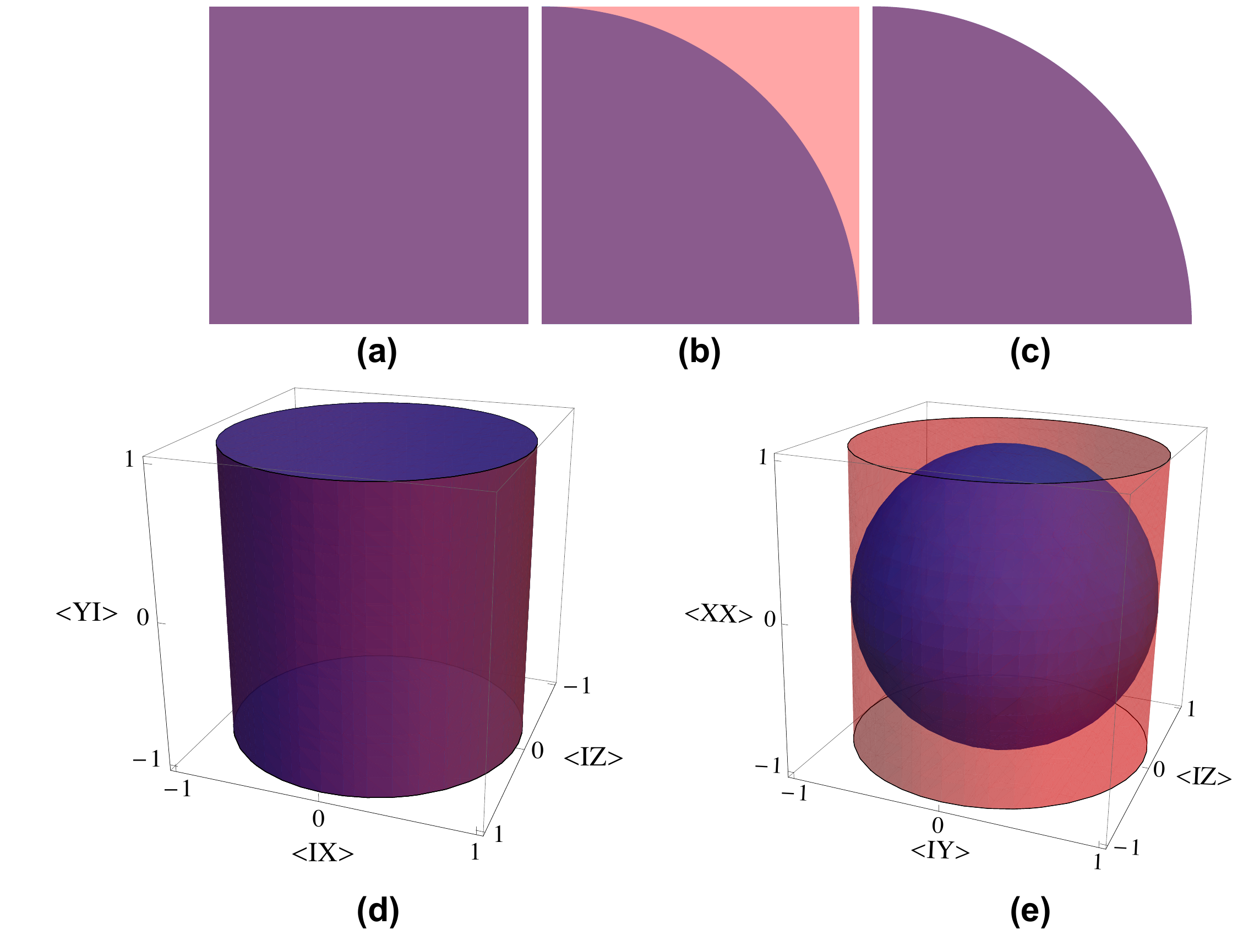}
\caption{\noindent
This figure present the types of correlations in two-point PDMs as 2-D projections onto the planes of $\{\left<\sigma_{A_1}\sigma_{B_1}\right>,\left<\sigma_{A_2}\sigma_{B_2}\right>\}$ in \textbf{(a)}, \textbf{(b)} and \textbf{(c)}. The sets of correlations are shown in the first quadrant. The full 2-D projection is generated in a symmetric manner about the origin. In \textbf{(d)} and \textbf{(e)} we give instances in the 3-D spaces corresponding to this 2-D projections. The red region highlights extra correlations attainable in a valid PDM compared to a valid density matrix.
\\
\textbf{(a)} Type a: $\left[\sigma_{A_1}\otimes\sigma_{B_1},\sigma_{A_2}\otimes\sigma_{B_2}\right]=0$. Temporal and spatial correlations both lie in the purple unit square.\\
\textbf{(b)} Type b: $\{\sigma_{A_1}\otimes\sigma_{B_1},\sigma_{A_2}\otimes\sigma_{B_2}\}=0$, and one out of the four operators is $\sigma_0$. Spatial correlations lie in the purple quarter unit circle, while temporal correlations lie in the unit square. The red region is allowed by valid PDMs but not density matrices.\\
\textbf{(c)} Type c: In all other cases, correlations are bounded by the purple quarter circle.\\
\textbf{(d)} An example of 3-D spaces corresponding to a combination of type a and type c 2-D projections. 
\\
\textbf{(e)} An example of 3-D spaces corresponding to a combination of type b and type c 2-D projections.
}\label{fig:2D}
\end{figure}

The remaining components of the two-point PDM concern all possible combinations of $\sigma_{A_1},\sigma_{A_2},\sigma_{B_1},\sigma_{B_2}\in\{\mathrm{I},\mathrm{X},\mathrm{Y},\mathrm{Z}\}$. We illustrate their geometry in FIG. \ref{fig:2D}. This completely characterizes the two-point spatial and temporal correlations for qubit systems. From FIG. \ref{fig:2D} it can be seen that the space of possible temporal correlations is strictly larger than the space of possible spatial correlations. These extra correlations cannot originate solely from spatially separated events, and hence are a signature of causal influence.

The presented geometrical structure can be applied to quantum causal inference \cite{ried2015quantum}. Given estimates for the expectation values of two-point correlations, one can identify the corresponding coordinates in the provided structure, and infer whether there exists a causal relationship between them. Our results also reveal pairs of temporal correlations that are statistically identical to entangled correlations in space.                  
An instance of this result is reflected in the violation of the temporal CHSH inequality \cite{brukner2004quantum}, which can be expressed entirely in terms of $\left<\sigma_1\sigma_1\right>$ and $\left<\sigma_2\sigma_2\right>$ correlations. The "inflated tetrahedron" imposes constraints in the space of all three Pauli correlations, hence serves as a stronger geometrical criterion for classifying quantum correlations and can act as a causal witness.\\

\textit{Acknowledgements}. The authors thank Artur Ekert and Otfried G\"uhne for insightful discussions, and Tommaso Demarie and Nana Liu for helpful comments on the manuscript. J.F.F. acknowledges support from the Air Force Office of Scientific Research under grant FA2386-15-1-4082. V.V. and R.P. thank the EPSRC (UK). V.V. thanks the Leverhulme Trust, the Oxford Martin School, and Wolfson College, University of Oxford. The authors acknowledge support from Singapore Ministry of Education. This material is based on research funded by the National Research Foundation of Singapore under NRF Award No. NRF-NRFF2013-01, NRF-NRFF2016-02 and the Competitive Research Programme (CRP Award No. NRF- CRP14-2014-02). This worked is also supported by the John Templeton Foundation Grant 53914 {\em ``Occam's Quantum Mechanical Razor: Can Quantum theory admit the Simplest Understanding of Reality?''} and the Foundational Questions Institute (FQXi).

\bibliographystyle{apsrev4-1}
\bibliography{causality}
\section*{Appendix}
Here we show the validity of Eq. (\ref{eq:PDM2}). We start by recalling the general expression for a PDM \cite{fitzsimons2015quantum}:
\begin{align}
R=\frac{1}{2^n}\sum\limits_{i_1=0}^{3}...\sum\limits_{i_n=0}^{3}\langle\{\sigma_{i_j}\}^n_{j=1}\rangle\bigotimes\limits_{j=1}^{n}\sigma_{i_j}.
\end{align}
We will also make use of the expression for the expectation value of the product of $n$ Pauli observables, which is given by 
\begin{align}
\langle\{\sigma_{i_j}\}^n_{j=1}\rangle = \Tr\left[\left(\bigotimes\limits_{j=1}^{n}\sigma_{i_j}\right)R\right].\label{correlation} 
\end{align}
In the case of two sequential events, $n=2$. Supposing the evolution between $t_A$ and $t_B$ is the identity, the only non-zero Pauli correlation functions are
\begin{align}
\left<\{\sigma_1,\sigma_1\}\right>&=\left<\{\sigma_2,\sigma_2\}\right>=\left<\{\sigma_3,\sigma_3\}\right>=\left<\{\sigma_0,\sigma_0\}\right>=1,\nonumber\\
\left<\{\sigma_0,\sigma_1\}\right>&=\left<\{\sigma_1,\sigma_0\}\right>=\left<\sigma_1\right>,\nonumber\\
\left<\{\sigma_0,\sigma_2\}\right>&=\left<\{\sigma_2,\sigma_0\}\right>=\left<\sigma_2\right>,\nonumber\\
\left<\{\sigma_0,\sigma_3\}\right>&=\left<\{\sigma_3,\sigma_0\}\right>=\left<\sigma_3\right>.
\end{align}

Here $\{...\}$ denotes sets of operators, not to be confused with anti-commutators.

On the other hand, a single-qubit density operator $\rho_A$ can be written as:
\begin{align}
\rho_{A}=\frac{1}{2}\left(\sigma_0+\left<\sigma_1\right>\sigma_1+\left<\sigma_2\right>\sigma_2+\left<\sigma_3\right>\sigma_3\right).
\end{align}
By comparing the coefficients of Pauli components, we can obtain the following useful form:
\begin{align}
R=\{\rho_A\otimes\frac{\mathrm{I}}{2},SWAP\}.
\end{align}
In a general setting, a channel that acts on the system in between the time instances $t_A$ and $t_B$
as a CPTP map $\varepsilon_{B|A}$ must be included. The map does not affect the observables at $t_A$, but introduces a transformation according to its adjoint map on the observables at $t_B$. Therefore the two-time PDM across such a channel can be written as
\begin{align}
R_{AB}=	(\mathcal{I}_A\otimes \varepsilon_{B|A})\{\rho_A\otimes\frac{\mathrm{I}}{2},SWAP\}.
\end{align}
By expanding the above equation into its Pauli components and substituting into \eqref{correlation}, we obtain
\begin{align}
\langle\sigma_k\sigma_k\rangle=\Tr\left[\langle\sigma_k\rangle_{\rho_A}\varepsilon_{B|A}(\sigma_0)\sigma_k+\varepsilon_{B|A}(\sigma_k)\sigma_k\right], \label{kk}
\end{align}
where $\langle\sigma_k\rangle_{\rho_A}$ denotes the expectation value of the $\sigma_k$ observable on the initial state $\rho_A$.

It was established in \cite{ruskai2002analysis} that the complete positivity requirement enables a particularly useful trigonometric parameterisation of the set of possible $\varepsilon_{B|A}$ in the Pauli basis \cite{king2001minimal}. Concretely, this set corresponds to the convex closure of the maps characterised by the following Kraus operators up to permutations among $\{\sigma_1,\sigma_2,\sigma_3\}$:
\begin{align}
K_+=\left[\cos\frac{v}{2}\cos\frac{u}{2}\right]\sigma_0+\left[\sin\frac{v}{2}\sin\frac{u}{2}\right]\sigma_3,\nonumber\\
K_-=\left[\sin\frac{v}{2}\cos\frac{u}{2}\right]\sigma_1-i\left[\cos\frac{v}{2}\sin\frac{u}{2}\right]\sigma_2,
\end{align}
where $v\in[0,\pi], u\in[0,2\pi]$. The above Kraus operators act on $\sigma_i$ as the following:
\begin{align}
	K_+^\dagger\sigma_0K_++K_-^\dagger\sigma_0K_-&=\sigma_0+\sin(u)\sin(v)\sigma_3,\nonumber\\
		K_+^\dagger\sigma_1K_++K_-^\dagger\sigma_1K_-&=\cos(u)\sigma_1,\nonumber\\ 
		K_+^\dagger\sigma_2K_++K_-^\dagger\sigma_2K_-&=\cos(v)\sigma_2,\nonumber\\ 
		K_+^\dagger\sigma_3K_++K_-^\dagger\sigma_3K_-&=\cos(u)\cos(v)\sigma_3.
\end{align}
By setting the $\rho_A=\ket{0}\bra{0}$ and apply the above Kraus operators, we obtain the parametric equations which characterise the convex set of possible correlation functions as followed:
\begin{align}
\langle\sigma_1\sigma_1\rangle&=\cos(u),\nonumber\\
\langle\sigma_2\sigma_2\rangle&=\cos(v),\nonumber\\
\langle\sigma_3\sigma_3\rangle&=\cos(u-v).\label{trig}
\end{align}
The first term in the trace of Eq. \eqref{kk} vanishes whenever either $\rho_A$ is maximally-mixed or $\varepsilon_{B|A}$ is a unital map, in which case the parametric equations reduce to 
\begin{align}
\langle\sigma_1\sigma_1\rangle&=\cos(u),\nonumber\\
\langle\sigma_2\sigma_2\rangle&=\cos(v),\nonumber\\
\langle\sigma_3\sigma_3\rangle&=\cos(u)\cos(v).
\end{align}
The above equations gives the extremal points $(1,1,1)$, $(1,-1,-1)$, $(-1,1,-1)$ and $(-1,-1,1)$, whose convex enclosure gives $\mathcal{T}_t$. We have presented the results in FIG. \ref{fig:StarInCube} assuming the initial state $\rho_A$ is maximally-mixed. However this result would be independent of $\rho_A$ if the channel $\varepsilon_{B|A}$ is unital, meaning $\varepsilon_{B|A}(\sigma_0)=\sigma_0$. This is because only non-unital maps act non-trivially on the local components $\sigma_k\otimes\sigma_0$ of the PDM, which leads to an augmented set of correlations. The choice of permutation among $\{\sigma_1,\sigma_2,\sigma_3\}$ is arbitrary and does not affect the resultant convex set enclosed by the parametric surface. 
\end{document}